\documentclass{article}
\usepackage{amssymb}
\usepackage{amsfonts}
\usepackage{amsmath}

\input{tcilatex}

\begin{document}

\title{First-order intertwining operators with position dependent mass and $%
\eta $-weak-pseudo-Hermiticity generators}
\author{Omar Mustafa$^{1}$ and S.Habib Mazharimousavi$^{2}$ \\
Department of Physics, Eastern Mediterranean University, \\
G Magusa, North Cyprus, Mersin 10,Turkey\\
$^{1}$E-mail: omar.mustafa@emu.edu.tr\\
$^{2}$E-mail: habib.mazhari@emu.edu.tr}
\maketitle

\begin{abstract}
A Hermitian and an anti-Hermitian first-order intertwining operators are
introduced and a class of $\eta $-weak-pseudo-Hermitian position-dependent
mass (PDM) Hamiltonians are constructed. A corresponding \emph{%
reference-target} $\eta $-weak-pseudo-Hermitian PDM -- Hamiltonians' map is
suggested. Some $\eta $-weak-pseudo-Hermitian $\mathcal{PT}$ -symmetric
Scarf II and periodic-type models are used as illustrative examples. \emph{%
Energy-levels crossing }and \emph{flown-away states} phenomena are reported
for the resulting Scarf II spectrum. Some of the corresponding $\eta $%
-weak-pseudo-Hermitian Scarf II- and periodic-type-isospectral models ($%
\mathcal{PT}$ -symmetric and non-$\mathcal{PT}$ -symmetric) are given as
products of the reference-target map.

\end{abstract}

\section{Introduction}

Recently, the growing interest in non-Hermitian Hamiltonians with real
spectra [1-9] was initiated by Bender's and Boettcher's [1] study of the
non-Hermitian Hamiltonian%
\begin{equation}
H=p^{2}+x^{2}\left( ix\right) ^{\nu };\text{ }\nu \geq 0.
\end{equation}%
They have observed that with properly defined boundary conditions the
spectrum of (1) is real, positive and discrete. It is, thereafter, concerted
that the reality of the spectrum of some non-Hermitian Hamiltonians might \
very well be attributed to their $\mathcal{PT}$ -symmetric settings (where $%
\mathcal{P}$ denotes parity and $\mathcal{T}$ mimics the time-reversal).
More specifically, if $\mathcal{PT}H\mathcal{PT}=H$ and if $\mathcal{PT}\Phi
\left( x\right) =\pm \Phi \left( x\right) $ the eigenvalues turn out to be
real. However, if the latter condition is not satisfied the eingenvalues
appear in complex-conjugate pairs (cf., e.g., Z. Ahmed in [1]). The
Hermiticity condition as a necessity for the reality of the spectrum is
relaxed therefore.

In a series of papers [3], Mostafazadeh has explored the necessary and
sufficient conditions for the reality of the spectra of more general
diagonalizable Hamiltonians (not necessarily restricted to Hermiticity
condition). \ He emphasized that $\mathcal{PT}$ -symmetric Hamiltonians form
a subclass of the so-called pseudo-Hermitian Hamiltonians [3,4]. That is, a
Hamiltonian $H$ is pseudo-Hermitian if it obeys the similarity transformation%
\begin{equation*}
\eta \,H\,\eta ^{-1}=H^{\dagger },
\end{equation*}%
where $\eta $ is a Hermitian invertible linear operator and $(^{\dagger })$
denotes the adjoint.

It is now a consensus that neither Hermiticity nor $\mathcal{PT}$ -symmetry
serve as necessary conditions for the reality of the spectrum of a quantum
Hamiltonian [3-10]. Yet, the existence of real eigenvalues is realized to be
associated\ with a non-Hermitian Hamiltonian provided that it is an $\eta $%
-pseudo-Hermitian:%
\begin{equation}
\eta \,H=H^{\dagger }\,\eta ,
\end{equation}%
with respect to the nontrivial "metric"operator $\eta =O^{\dagger }O$, for
some linear invertible operator $O:\mathcal{H}{\small \rightarrow }\mathcal{H%
}$ ($\mathcal{H}$ is the Hilbert space).

However, under some rather mild assumptions, we may even relax $H$ to be an $%
\eta $-weak-pseudo-Hermitian by not restricting $\eta $ to be Hermitian
(cf., e.g., Bagchi and Quesne [5]), and linear and/or invertible (cf., e.g.,
Solombrino [6], Fityo [7], and Mustafa and Mazharimousavi [8]). A
classification of this sort is necessary to avoid contradictions with the
well established $\eta $-pseudo-Hermiticity theorems by Mostafazadeh [3].

Without enforcing invertibility in the process, we have very recently
introduced a class of spherically symmetric non-Hermitian Hamiltonians and
their $\eta $-weak-pseudo-Hermiticity generators [9], where a generalization
beyond the nodeless 1D states was proposed. We have used the same recipe and
extend it for a class of $\eta $-weak-pseudo-Hermitian Hamiltonians for
quantum particles endowed with position-dependent mass (PDM) [8]. We were
inspired by the fact that a quantum particle endowed with position-dependent
mass, $M\left( x\right) =m_{\circ }m\left( x\right) $, constitutes a useful
model for the study of many physical problems [10-14]. For example, they are
used in the energy density many-body problem [12], in the determination of
the electronic properties of semiconductors [13] and quantum dots [14], etc.

The above forms the stimuli and inspiration of this paper which is organized
as follows. In section 2, we introduce two first-order intertwining
PDM-differential operators to construct a class of $\eta $%
-weak-pseudo-Hermitian PDM-Hamiltonians. A \emph{reference} and \emph{target}
$\eta $-weak-pseudo-Hermitian PDM-Hamiltonians' map is also introduced in
the same section. In section 3, an $\eta $-weak-pseudo-Hermitian
complexified $\mathcal{PT}$ -symmetric Scarf II model is used as an
illustrative \emph{reference} model. Two $\eta $-weak-pseudo-Hermitian Scarf
II-isospectral models are given as products of the \emph{reference-target}
map (for the sake of completeness of this work), in the same section. An $%
\eta $\emph{-}weak-pseudo-Hermitian $PT$-symmetric reference periodic-type
model is given in section 4, along with one of its isospectral \emph{%
reference-target} map descendants. We conclude in section 5.

\section{First-order intertwining PDM-operator $\protect\eta $ and $\protect%
\eta $-weak-pseudo-Hermiticity}

For the sake of completeness and to keep the current study minimally
self-contained, we recollect that one may avoid invertibility as a necessity
for $\eta $ and consider a non-Hermitian $\eta $-weak-pseudo-Hermitian
Hamiltonian $H$ satisfying the intertwining relation $\eta H=H^{\dagger
}\,\eta $. Then, if $\eta _{1}$ is Hermitian, it is easy to show that $%
\left( \eta _{1}H\right) $ is also Hermitian (cf.,e.g., [7]). On the other
hand, if $\eta _{2}$ is anti-Hermitian such that $\eta _{2}=i\eta _{1}$,
then $\left( \eta _{2}H\right) $ is anti-Hermitian too$.$Yet, in both cases,
the Hamiltonian $H$ may very well be classified as an $\eta $%
-weak-pseudo-Hermitian. The proof of the reality of the spectrum in both
cases is straightforward:

Let $\psi _{n}\left( x\right) $ and $E_{n}$ be the eigenfunctions and
eigenvalues of $H$, respectively. Then, the Hermiticity of $\left( \eta
_{1}H\right) $ implies that 
\begin{equation*}
\langle \psi _{n}\mid \eta _{1}H\mid \psi _{n}\rangle =E_{n}\langle \psi
_{n}\mid \eta _{1}\mid \psi _{n}\rangle ,
\end{equation*}%
where%
\begin{equation*}
\langle \psi _{n}\mid \eta _{1}H\mid \psi _{n}\rangle \in 
\mathbb{R}
\ni \langle \psi _{n}\mid \eta _{1}\mid \psi _{n}\rangle \Longrightarrow
E_{n}\in 
\mathbb{R}
.
\end{equation*}%
Next, the anti-Hermiticity of $\left( \eta _{2}H\right) $ yields (with $\eta
_{2}=i\eta _{1}$)%
\begin{equation*}
\langle \psi _{n}\mid \eta _{2}H\mid \psi _{n}\rangle =E_{n}\langle \psi
_{n}\mid \eta _{2}\mid \psi _{n}\rangle \text{ }\Rightarrow i\langle \psi
_{n}\mid \eta _{1}H\mid \psi _{n}\rangle =iE_{n}\langle \psi _{n}\mid \eta
_{1}\mid \psi _{n}\rangle .
\end{equation*}
Then, $\langle \psi _{n}\mid \eta _{2}H\mid \psi _{n}\rangle $ and $\langle
\psi _{n}\mid \eta _{2}\mid \psi _{n}\rangle $ are identically pure
imaginary resulting in $E_{n}\in 
\mathbb{R}
$.

Hence, the reality of the spectrum for both cases is secured provided that $%
\langle \psi _{n}\mid \eta _{i}\mid \psi _{n}\rangle \neq 0$, otherwise the
reality/imaginary nature of $E_{n}$ remains undetermined. In the forthcoming
developments we shall focus only on the Hamiltonians satisfying the above
mentioned properties.

\subsection{Consequences of Hermiticity and anti-Hermiticity on a
first-order intertwining PDM-operator}

A first-order differential operator $\eta _{1}$ with position-dependent mass
settings:%
\begin{equation}
\eta _{1}=-i\,\left[ \mu \left( x\right) \,\partial _{x}+G_{1}\left(
x\right) \right] +F_{1}\left( x\right)
\end{equation}%
is Hermitian if and only \ if $G_{1}\left( x\right) =$ $\mu ^{\prime }\left(
x\right) /2,$ where $\mu \left( x\right) =1/\sqrt{M\left( x\right) }$, $%
M\left( x\right) =m_{\circ }m\left( x\right) ,$ and prime denotes derivative
with respect to $%
\mathbb{R}
\ni x\in \left( -\infty ,\infty \right) $. On the other hand, a first-order
PDM-differential operator%
\begin{equation}
\eta _{2}=\mu \left( x\right) \,\partial _{x}+G_{2}\left( x\right)
+i\,F_{2}\left( x\right)
\end{equation}%
is anti-Hermitian if and only if $G_{2}\left( x\right) =$ $\mu ^{\prime
}\left( x\right) /2$. It is obvious that $\eta _{2}=i\eta _{1}$. Moreover, a
symmetry ordering recipe of the momentum and position dependent mass would
imply a non-Hermitian Schr\"{o}dinger Hamiltonian%
\begin{equation}
H=-\partial _{x}\left( \frac{1}{M\left( x\right) }\right) \partial
_{x}+V\left( x\right) +iW\left( x\right) ,
\end{equation}%
where $\hbar =2m_{\circ }=1$ and $\alpha =\gamma =0$ and $\beta =-1$ in
equation (6) of von Roos in [10].

Within the above settings, the intertwining relation%
\begin{equation}
\eta _{j}\,H=H^{\dagger }\,\eta _{j}\text{ \ }\,;\text{ }j=1,2\text{\ ,\ }
\end{equation}%
(where $F_{j}\left( x\right) ,G_{j}\left( x\right) ,V\left( x\right)
,W\left( x\right) \in 
\mathbb{R}
$) would imply%
\begin{equation}
2\mu \left( x\right) ^{2}F_{j}^{\prime }\left( x\right) +2\mu \left(
x\right) W\left( x\right) =0,
\end{equation}%
\begin{equation}
2\mu \left( x\right) \mu ^{\prime }\left( x\right) F_{j}^{\prime }\left(
x\right) +\mu ^{\prime }\left( x\right) W\left( x\right) +\mu \left(
x\right) W\,^{\prime }\left( x\right) +\mu \left( x\right) ^{2}F_{j}^{\prime
\prime }\left( x\right) =0,
\end{equation}%
\begin{equation}
-\mu \left( x\right) V\,^{\prime }\left( x\right) -\frac{1}{2}\mu \left(
x\right) ^{2}\mu ^{\prime \prime \prime }\left( x\right) +2F_{j}\left(
x\right) W\left( x\right) -\mu \left( x\right) \mu ^{\prime }\left( x\right)
\mu ^{\prime \prime }\left( x\right) =0.
\end{equation}%
Which in turn yields%
\begin{equation}
W\left( x\right) =-\mu \left( x\right) F_{j}^{\prime }\left( x\right) ,
\end{equation}%
\begin{equation}
V\left( x\right) =-F_{j}\left( x\right) ^{2}-\frac{1}{2}\mu \left( x\right)
\mu ^{\prime \prime }\left( x\right) -\frac{1}{4}\mu ^{\prime }\left(
x\right) ^{2}+\alpha _{\circ },
\end{equation}%
where $\alpha _{\circ }\in $ $%
\mathbb{R}
$ is an integration constant. Then our $\eta $-weak-pseudo-Hermitian
PDM-Hamiltonian (5) reads%
\begin{equation}
H=-\mu \left( x\right) ^{2}\partial _{x}^{2}-2\mu \left( x\right) \mu
^{\prime }\left( x\right) \partial _{x}+\tilde{V}_{j}\left( x\right) ,
\end{equation}%
where%
\begin{equation}
\tilde{V}_{j}\left( x\right) =-F_{j}\left( x\right) ^{2}-\frac{1}{2}\mu
\left( x\right) \mu ^{\prime \prime }\left( x\right) -\frac{1}{4}\mu
^{\prime }\left( x\right) ^{2}+\alpha _{\circ }-i\mu \left( x\right)
F_{j}^{\prime }\left( x\right) ,
\end{equation}%
and $\alpha _{\circ }\in $ $%
\mathbb{R}
$ may very well serve for some feasible exactly solvable $\eta $%
-weak-pseudo-Hermitian Hamiltonian models. Obviously, $F_{j}\left( x\right) $%
's are our generating functions for some $\eta $-weak-pseudo-Hermitian
Hamiltonians in (12).

\subsection{Corresponding $\protect\eta $-weak-pseudo-Hermitian
Hamiltonians' \emph{reference-target} map}

In this section we invest our experience in the
point-canonical-transformation [8,9,11] and consider our non-Hermitian $\eta 
$-weak-pseudo-Hermitian Hamiltonian $H$, in (12), in \ the one-dimensional
Schr\"{o}dinger equation:%
\begin{equation}
\left( H-E\,\right) \psi \left( x\right) =0,
\end{equation}%
to construct the so-called \emph{reference/old} and \emph{target/new}
non-Hermitian $\eta $-weak-pseudo-Hermitian Hamiltonians' map.

A substitution of the form 
\begin{equation*}
\psi \left( x\right) \,=\varphi \left( q\left( x\right) \right) /\sqrt{\mu
\left( x\right) }
\end{equation*}%
would result in the so-called \emph{target/new} Schr\"{o}dinger equation%
\emph{\ } 
\begin{gather}
\mu \left( x\right) ^{2}\left[ q^{\prime }\left( x\right) \right]
^{2}\,\partial _{q}^{2}\varphi \left( q\right) +\mu \left( x\right) \left[
\mu \left( x\right) q^{\prime }\left( x\right) \right] ^{\prime }\,\partial
_{q}\varphi \left( q\right)   \notag \\
+\left[ F_{j}\left( x\right) ^{2}+i\mu \left( x\right) F_{j}^{\prime }\left(
x\right) +E-\alpha _{\circ }\right] \varphi \left( q\right) =0.
\end{gather}%
Next, one may avoid the first-order derivative of $\varphi \left( q\right) $
to come out with the traditional form of the one-dimensional Schr\"{o}dinger
equation and substitute 
\begin{equation*}
q^{\prime }\left( x\right) =1/\mu \left( x\right) 
\end{equation*}%
to obtain%
\begin{equation}
-\partial _{q}^{2}\varphi \left( q\right) +\left[ \alpha _{\circ
}-F_{j}\left( x\left( q\right) \right) ^{2}-i\mu \left( x\right)
F_{j}^{\prime }\left( x\right) -E\right] \varphi \left( q\right) =0.
\end{equation}%
This in turn, with%
\begin{equation*}
\frac{dF_{j}\left( x\right) }{dx}=\frac{dF_{j}\left( q\left( x\right)
\right) }{dx}=\frac{dq\left( x\right) }{dx}\frac{dF_{j}\left( q\right) }{dq}=%
\frac{1}{\mu \left( x\right) }\frac{dF_{j}\left( q\right) }{dq},
\end{equation*}
collapses into the \emph{reference/old} Schr\"{o}dinger equation%
\begin{equation}
-\partial _{q}^{2}\varphi \left( q\right) +\left[ \tilde{V}_{eff}\left(
q\right) -E\right] \varphi \left( q\right) =0,
\end{equation}%
where%
\begin{equation}
\tilde{V}_{eff}\left( q\right) =\alpha _{\circ }-F_{j}\left( q\right)
^{2}-iF_{j}^{\prime }\left( q\right) .
\end{equation}%
The \emph{reference-target}\ map is therefore complete and an explicit
correspondence between two bound state problems is obtained. That is, one
needs to find the exact, quasi-exact, or conditionally-exact solution
(eigenvalues and eigenfunctions) for the \emph{reference/old} Schr\"{o}%
dinger equation (17) and map it into the \emph{target/new} Schr\"{o}dinger
equation (14), where $H$ is defined in (12).

\section{An $\protect\eta $\emph{-}weak-pseudo-Hermitian $PT$-symmetric
reference Scarf II model}

A complexified $PT$-symmetric Scarf II potential%
\begin{equation}
V\left( x\right) =-V_{1}\func{sech}^{2}q-iV_{2}\func{sech}q\,\tanh q;\text{ }%
V_{1}>0,\text{ }V_{2}\neq 0,V_{1},V_{2}\in 
\mathbb{R}
\text{, }
\end{equation}%
is studied by Bagchi and Quesne [15] using complex Lie algebras (sl$\left( 2,%
\mathbb{C}
\right) $ in particular). Therein, exact real eigenvalues are reported:%
\begin{equation}
E_{n,\epsilon }=-\left[ \frac{1}{2}\left( \sqrt{V_{1}+\frac{1}{4}+\left\vert
V_{2}\right\vert }+\epsilon \sqrt{V_{1}+\frac{1}{4}-\left\vert
V_{2}\right\vert }\right) -n-\frac{1}{2}\right] ^{2},\text{ }\epsilon =\pm 1,
\end{equation}%
where 
\begin{equation*}
n=0,1,2,\cdots <\frac{1}{2}\left( \sqrt{V_{1}+\frac{1}{4}+\left\vert
V_{2}\right\vert }+\epsilon \sqrt{V_{1}+\frac{1}{4}-\left\vert
V_{2}\right\vert }-1\right) .
\end{equation*}

On the other hand, an $\eta $\emph{-}weak-pseudo-Herrmiticity generator of
the form 
\begin{equation}
F_{j}\left( q\right) =-V_{2}\func{sech}q\Longrightarrow F_{j}^{\prime
}\left( q\right) =V_{2}\func{sech}q\tanh q
\end{equation}%
would imply (with $\alpha _{\circ }=0$)%
\begin{equation}
\tilde{V}_{eff}\left( q\right) =-V_{2}^{2}\func{sech}^{2}q-iV_{2}\func{sech}%
q\tanh q.
\end{equation}%
Then the eigenvalues in (20), with $V_{1}=V_{2}^{2}$, read 
\begin{equation}
E_{n,\epsilon }=-\left[ \frac{1}{2}\left( \sqrt{V_{2}^{2}+\frac{1}{4}%
+\left\vert V_{2}\right\vert }+\epsilon \sqrt{V_{2}^{2}+\frac{1}{4}%
-\left\vert V_{2}\right\vert }\right) -n-\frac{1}{2}\right] ^{2},\text{ }%
\epsilon =\pm 1,
\end{equation}%
which collapses, with $\left\vert V_{2}\right\vert >1/2$ and $\epsilon =+1,$
into%
\begin{equation}
E_{n,\epsilon =+1}=-\left[ \left\vert V_{2}\right\vert -n-\frac{1}{2}\right]
^{2}\text{ };\text{ \ }n=0,1,2,\cdots ,n_{\max }<\left( \left\vert
V_{2}\right\vert -1/2\right) .
\end{equation}%
However, for $\left\vert V_{2}\right\vert <1/2$ and $\epsilon =\pm 1,$ and
for $\left\vert V_{2}\right\vert >1/2$ and $\epsilon =-1$ one would get
empty sets of energy eigenvalues.

Nevertheless, it is obvious that the phenomenon of \emph{energy-levels
crossing} is manifested (cf., e.g., Mustafa and Znojil [2]) in the current $%
PT$-symmetric Scarf II energy spectrum in (24). Such a phenomenon occurs
when a state $E_{n_{1}}$ at $\left\vert V_{2}\right\vert =V_{2,1}$ and a
state $E_{n_{2}}$ at $\left\vert V_{2}\right\vert =V_{2,2}$ have the same
energy eigenvalues. In this case%
\begin{equation*}
E_{n_{1}}\left( \left\vert V_{2}\right\vert =V_{2,1}\right) =E_{n_{2}}\left(
\left\vert V_{2}\right\vert =V_{2,2}\right) ;\ n_{2}>n_{1},
\end{equation*}%
and hence the energy-level crossing occurs when%
\begin{equation*}
V_{2,2}-\Delta n=V_{2,1};\Delta n=n_{2}-n_{1}>0.
\end{equation*}%
It might also be interesting to observe that the so called \emph{flown away}
states phenomenon (cf., e.g., Mustafa and Znojil [2]) is feasible when $%
\left\vert V_{2}\right\vert >>n$. In this case, such states \emph{fly away}
and disappear from the spectrum.

\subsection{Corresponding Scarf II-isospectral $\protect\eta $\emph{-}%
weak-pseudo-Hermitian models}

Let us start with a general case and consider $q\left( x\right) =\pm \ln
f\left( x\right) ;$ $f\left( x\right) \in 
\mathbb{R}
$ to imply%
\begin{equation}
M\left( x\right) =\left[ \partial _{x}\ln .f\left( x\right) \right]
^{2}\Longrightarrow f\left( x\right) =\exp \left( \pm \dint^{x}\sqrt{M\left(
z\right) }dz\right)
\end{equation}%
and%
\begin{equation}
\tilde{V}_{eff}\left( x,q=\pm \ln f\left( x\right) \right) =-4V_{2}^{2}\frac{%
\,f\left( x\right) ^{2}}{\left( f\left( x\right) ^{2}+1\right) ^{2}}\mp
2iV_{2}\frac{f\left( x\right) \left( f\left( x\right) ^{2}-1\right) }{\left(
f\left( x\right) ^{2}+1\right) ^{2}}.
\end{equation}%
Then, a class of Scarf II-isospectral $\eta $\emph{-}weak-pseudo-Hermitian
models is now generated and $\,f\left( x\right) $ may very well be
considered as a Scarf II-isospectral PDM $\eta $\emph{-}%
weak-pseudo-Hermiticity generator. Of course, as long as a chosen $f\left(
x\right) $ generates a physically acceptable position dependent mass
function $M\left( x\right) $.

Under such settings, one may conclude that the number of the related Scarf
II-isospectral $\eta $\emph{-}weak-pseudo-Hermitian models is large. we only
choose two (in one) illustrative models:

An $f\left( x\right) =\left( x^{2}+1\right) $ would lead to%
\begin{equation}
q\left( x\right) =\ln \left( x^{2}+1\right) =\int^{x}\sqrt{M\left( z\right) }%
dz\Longrightarrow M\left( x\right) =\frac{4x^{2}}{\left( x^{2}+1\right) ^{2}}
\end{equation}%
and%
\begin{equation}
\tilde{V}_{eff}\left( x\right) =-4V_{2}^{2}\frac{\left( x^{2}+1\right) ^{2}}{%
\left( \left( x^{2}+1\right) ^{2}+1\right) ^{2}}-2iV_{2}\frac{\left(
x^{2}+1\right) \left( \left( x^{2}+1\right) ^{2}-1\right) }{\left( \left(
x^{2}+1\right) ^{2}+1\right) ^{2}},
\end{equation}%
which is a non-singular non-$\mathcal{PT}$-symmetric $\eta $\emph{-}%
weak-pseudo-Hermitian model.

\section{An $\protect\eta $\emph{-}weak-pseudo-Hermitian $PT$-symmetric
reference periodic-type model}

An $\eta $\emph{-}weak-pseudo-Herrmiticity generator of the form

\begin{equation}
F_{j}(q)=-\frac{4}{3\cos ^{2}q-4}-\frac{5}{4}
\end{equation}%
would imply (with $\alpha _{\circ }=0$) an effective periodic-type $PT$%
-symmetric reference potential of the form%
\begin{equation}
\tilde{V}_{eff}(q)=\frac{1}{9}\frac{\left( -30\cos ^{2}q+24\right) }{(\cos
^{2}q-\frac{4}{3})^{2}}+\frac{4i\sin 2q}{3(\cos ^{2}q-\frac{4}{3})^{2}}-%
\frac{25}{16}.
\end{equation}%
In a straightforward manner one may show that%
\begin{equation}
\tilde{V}_{eff}(q)=-\frac{6}{(\cos q+2i\sin q)^{2}}-\frac{25}{16}.
\end{equation}%
Such an effective potential represents a "shifted by $-\frac{25}{16}$"
Samsonov's and Roy's [2] periodic potential model. The solution of which is
reported for the interval $q\in \left( -\pi ,\pi \right) $ with the boundary
conditions $\varphi \left( -\pi \right) =\varphi \left( \pi \right) =0$ as%
\begin{eqnarray}
\varphi \left( q\right) &=&\left\{ \left[ \left( 16-n^{2}\right) \cos
q-2i\left( n^{2}-4\right) \sin q\right] \sin \left[ \frac{n}{2}\left( \pi
+q\right) \right] \right.  \notag \\
&&\left. -6n\sin q\cos \left[ \frac{n}{2}\left( \pi +q\right) \right]
\right\} (\cos q+2i\sin q)^{-1}
\end{eqnarray}%
and%
\begin{equation}
E_{n}=\frac{n^{2}}{4}-\frac{25}{16}\text{ };\text{ \ }n=1,3,4,5,\cdots
\end{equation}%
with a missing $n=2$ state (the details of which can be found in Samsonov
and Roy [2])

\subsection{Corresponding periodic-type-isospectral $\protect\eta $\emph{-}%
weak-pseudo-Hermitian models}

One of the choices for $q\left( x\right) $ is the class of models descending
from $q\left( x\right) =\arctan g\left( x\right) ;$ $g\left( x\right) \in 
\mathbb{R}
$ to imply%
\begin{equation}
M\left( x\right) =\left[ \frac{\partial _{x}.g\left( x\right) }{1+g\left(
x\right) ^{2}}\right] ^{2},
\end{equation}%
and%
\begin{equation}
\tilde{V}_{eff}\left( x\right) =-\frac{6\left( g\left( x\right)
^{2}+1\right) }{(1+2ig\left( x\right) )^{2}}-\frac{25}{16}.
\end{equation}%
Then, a class of periodic-type isospectral non-$\mathcal{PT}$-symmetric (for
even $g\left( x\right) $ and $\mathcal{PT}$-symmetric for odd $g\left(
x\right) $) $\eta $\emph{-}weak-pseudo-Hermitian models (not necessarily
periodic-type) is generated, where $\,g\left( x\right) $ is now a
periodic-type isospectral PDM $\eta $\emph{-}weak-pseudo-Hermiticity
generator. Again, as long as a chosen $g\left( x\right) $ generates a
physically acceptable position dependent mass function $M\left( x\right) $.

To illustrate the process, let us take $g\left( x\right) =x$ then $M\left(
x\right) =\left( 1+x^{2}\right) ^{-2}$ and%
\begin{equation}
\tilde{V}_{eff}\left( x\right) =-\frac{6\left( x^{2}+1\right) }{(1+2ix)^{2}}-%
\frac{25}{16}\text{ };\text{ \ }%
\mathbb{R}
\ni x\in \left( -\infty ,\infty \right) .
\end{equation}%
Next, one may choose to work the other way around and start with $M\left(
x\right) =\left( 1+x^{2}\right) ^{-2}$ to come out with the same $\tilde{V}%
_{eff}\left( x\right) $ in (36).

\section{Conclusion}

In this work, we have introduced two first-order intertwining
PDM-differential operators (a Hermitian $\eta _{1}$ and an anti-Hermitian $%
\eta _{2}=i\eta _{1}$) and constructed a class of $\eta $%
-weak-pseudo-Hermitian PDM-Hamiltonians. We have observed that the
Hermiticity of $\left( \eta _{1}H\right) $ and the anti-Hermiticity of $%
\left( \eta _{2}H\right) $ leaves the form of the resulting $\eta $%
-weak-pseudo-Hermitian PDM-Hamiltonian invariant (cf. (16) and/or
equivalently (18)) regardless of the Hermiticity nature of our intertwining
first-order differential operators $\eta _{1}$ and\ $\eta _{2}$. Moreover,
we have used a Liouvilean-type change of variables, $q^{\prime }\left(
x\right) =1/\mu \left( x\right) $, and constructed the corresponding \emph{%
reference-target} $\eta $-weak-pseudo-Hermitian PDM - Hamiltonians' map.
Hence, an explicit correspondence between two bound-state problems is
obtained.

Within the setting of such \emph{reference-target} map, we have generated a
complexified $\mathcal{PT}$ -symmetric Scarf II model and
reported/fine-tuned its exact eigenvalues along with two $\eta $%
-weak-pseudo-Hermitian Scarf II-isospectral models. We have also generated
an $\eta $\emph{-}weak-pseudo-Hermitian $PT$-symmetric reference
periodic-type model and reported one of its isospectral \emph{%
reference-target} map descendants. Nevertheless, one should be reminded that
the number of the isospectral \emph{reference-target} map descendants is not
only limited to one and/or two but rather remains elusive and unexplored as
yet.

Under the parametric settings generated for our complexified $\mathcal{PT}$
-symmetric Scarf II model, we have reported and discussed the phenomenon of 
\emph{energy-levels crossing} and the feasible manifestation of \emph{flown
away states} phenomenon. The reader may refer to Mustafa and Znojil [2] for
more details on these phenomena. Nevertheless, more comprehensive details on
the energy levels-crossing phenomenon the reader may wish to refer to, e.g.,
Guida et al. [17] who studied the energy-levels crossing of fermionic
systems in Instanton-Anti-Instanton valley, Nishino and Deguchi [18] who
discussed energy-levels crossings in the one dimensional Hubbard model, and
Bhattacharya and Raman [19] who elaborated on the detection of energy-levels
crossings and presented an algebraic method of finding such crossings.

However, in the process of testing a complexified non-$\mathcal{PT}$
-symmetric Morse model, we observed that a generating function of the form $%
F_{j}\left( q\right) =\eta e^{-q}$ would lead to%
\begin{equation}
V_{eff}\left( q\right) =-\eta ^{2}e^{-2q}+i\eta e^{-q}.
\end{equation}%
The bound-state solutions of which form an empty set of eigenvalues (cf.,
e.g., Bagchi and Quesne [15] and Ahmed [16]). Although the potential in (26)
remains an $\eta $-weak-pseudo-Hermitian, it rather represents an
unfortunate \emph{reference} model and consequently leading to a set of
unfortunate isospectral $\eta $-weak-pseudo-Hermitian \emph{target}
Hamiltonians.

\textbf{Aknowledgement: }We would like to thank the referee for the
tremendous effort and valuable suggestions.

\vspace{0pt}

\end{document}